\documentclass{ws-procs975x65}

\begin{document}

\title{BLACK HOLES AND QUASIBLACK HOLES IN EINSTEIN-MAXWELL THEORY}

\author{REINHARD MEINEL, MARTIN BREITHAUPT and YU-CHUN LIU}

\address{University of Jena, Theoretisch-Physikalisches Institut,\\
Max-Wien-Platz 1, 07743 Jena, Germany\\
\email{meinel@tpi.uni-jena.de}}

\begin{abstract}
Continuous sequences of asymptotically flat solutions to the Einstein-Maxwell
equations describing regular equilibrium 
configurations of ordinary matter can reach a black hole limit. For a distant
observer, the spacetime becomes more and more indistinguishable from the metric
of an extreme Kerr-Newman black hole outside the horizon when approaching the
limit. From an internal perspective, a still regular but non-asymptotically
flat spacetime with the extreme Kerr-Newman near-horizon geometry at spatial
infinity forms at the limit. Interesting special cases are sequences of
Papapetrou-Majumdar distributions of electrically counterpoised dust leading to
extreme Reissner-Nordstr\"om black holes and sequences of rotating uncharged
fluid bodies leading to extreme Kerr black holes.
\end{abstract}

\bodymatter

\section{Introduction}
Spherically symmetric (uncharged) perfect fluid bodies in equilibrium cannot reach a black
hole limit parametrically. Buchdahl has shown\cite{buchdahl} that the
Schwarzschild coordinate radius of such a fluid 
ball is always greater than $(9/8)\,2M$, where $M$ denotes the gravitational
mass\footnote{We use units in which $G=c=1$.}. Correspondingly, the relative
gravitational redshift $z$ of photons emitted from the fluid's surface and
received at infinity is bounded by $z<2$. 

With rotation and/or electric charge a (quasi) black hole limit of regular
equilibrium configurations is possible. 
In this parameter limit, if discussed from an external perspective, the
spacetime of an extreme Kerr-Newman black hole\footnote{For a constructive
proof of the Kerr-Newman black hole uniqueness including the extreme case, see
Ref.~\refcite{m12}.} outside the horizon forms. From
an internal perspective, a still regular but non-asymptotically flat
spacetime with the extreme Kerr-Newman near-horizon geometry at spatial infinity
results. The first example was given with the extreme relativistic limit of a
uniformly rotating disc of dust\cite{bw, nm}, for details see also 
Refs.~\refcite{m02, rfe, klm}. More examples can be found in 
Refs.~\refcite{bfm, lw, akm, fha, lpa, lemoszanchin}. The simplest way
to construct such ``quasiblack holes'' is to use the parametric compression of
static configurations of ``electrically counterpoised dust (ECD)'', also called
``Bonnor stars''\cite{bonnorwick, b98, b99, lemoswein, lz, b10, mh}. 

In the following, we will touch on some general aspects of the black hole limit
in the two special 
cases: (i) perfect fluid configurations (angular momentum but no charge) leading
to extreme Kerr black holes and (ii) ECD configurations (electric charge but no
angular momentum) leading to extreme Reissner-Nordstr\"om black holes. We
conclude with some comments on rotating discs of charged dust. This model
provides a continuous connection between the two cases mentioned above.

\section{Black hole limit of rotating fluid bodies in equilibrium}
A black hole limit of a stationary and axisymmetric, uniformly rotating
perfect fluid body occurs if and only if $M-2\Omega J \to 0$ holds at
the limit\cite{m06}, where $\Omega$ and $J$ denote angular velocity and angular
momentum. Consequently, from an external perspective, the spacetime of an
extreme Kerr black hole outside the horizon forms. A rigorous proof for the
existence of such a limit was given with the analytic solution to the disc of
dust problem\cite{nm, m02}. Convincing numerical evidence for such a limit was
also found for fluid ring solutions with various equations of
state\cite{akm,fha,lpa}. 

\section{Black hole limit of electrically counterpoised dust configurations}
The black hole limit of ECD configurations can be treated in quite a general
manner, by applying a simple scaling transformation to any
asymptotically flat Papapetrou-Majumdar solution of the Einstein-Maxwell
equations, corresponding to a localized ECD distribution\cite{mh}. It is
interesting to note that although the ``outer world'' at the limit is always
given by the spherically symmetric extreme Reissner-Nordstr\"om metric outside
the horizon, the ``inner world'' does not need to show any spatial symmetry in
general.  

\section{Rotating discs of charged dust}
The problem defining the global solution to the Einstein-Maxwell equations 
corresponding  to a uniformly rotating (infinitesimally thin) disc of 
electrically charged dust with constant specific charge $\epsilon$ 
($0\le \epsilon \le 1$, we are using Gauss units) can be formulated as a
boundary value problem for the 
Ernst equations\cite{e}
\begin{equation*}
f\,\Delta\, {\mathcal E}=(\nabla {\mathcal E} +
2\bar{\Phi}\nabla \Phi)\cdot\nabla {\mathcal E}\, , \quad
f\,\Delta\, \Phi=(\nabla {\mathcal E} +2\bar{\Phi}\nabla \Phi)\cdot\nabla \Phi
\end{equation*}
\begin{equation*}
\mbox{with} \quad f\equiv\Re\, {\mathcal E} + |\Phi|^2\, , \quad 
\Delta=\frac{\partial^2}{\partial\rho^2}+\frac{1}{\rho}\frac{\partial}{\partial\rho}
+\frac{\partial^2}{\partial\zeta^2} \, , \quad \nabla
=(\frac{\partial}{\partial\rho},\frac{\partial}{\partial\zeta})\, .
\end{equation*}
The boundary conditions on the disc ($\zeta=0$, $0 \le \rho \le \rho_0$)
take a simple form in the corotating frame of reference (indicated by primes):
\begin{equation}\label{BC}
\rho_0\frac{\partial}{\partial \rho}\left(\sqrt{f'}+
\epsilon\, \Re\,\Phi'\right) = 0\, , \;
\rho_0\frac{\partial}{\partial \zeta}\left(\Re\, \Phi' + 
\epsilon\,\sqrt{f'}\right) = 0\, , \;
\Im\,{\mathcal E'}=0\, , \; \Im\,\Phi'=0\, .
\end{equation}
They have to be complemented by the asymptotic flatness conditions
$\mathcal E \to 1$, $\Phi \to 0$ as $\rho^2+\zeta^2\to\infty$
and a regularity condition at the rim of the disc.
Except from a scaling parameter ($\rho_0$ or $M$), the solution can be discussed in 
terms of its dependence on the two parameters $\epsilon$ and
$\gamma\equiv 1-\sqrt{f(\rho=0,\,\zeta=0)}$ with $0< \gamma\le 1$. The limit $\epsilon\to 0$ leads back to 
the disc of dust solution discussed above. For $\epsilon\to 1$, one obtains a 
special (non-rotating) ECD configuration. The Newtonian limit is characterized by 
$\gamma \ll 1$, whereas $\gamma\to 1$ is expected to lead to the black hole
limit. The gravitational mass of the disc can, in general, be expressed as
\begin{equation}\label{M}
M=2\Omega J + (1-\gamma+\epsilon\, \alpha_0)M_0\, ,
\end{equation}
where $\alpha_0\equiv\Re\,\Phi(\rho=0,\,\zeta=0)$ and $M_0$ denotes the baryonic 
mass. Accordingly, the electric charge of the disc is given by $Q=\epsilon\,M_0$. The
factor $1-\gamma+\epsilon\, \alpha_0$ on the right-hand side of Eq.~\eqref{M} is equal to
the constant value of $\sqrt{f'}+\epsilon\, \Re\, \Phi'$ on the disc, cf.~the first 
boundary condition in \eqref{BC}. At the limit $\gamma\to 1$, we expect to
obtain $f'=0$ and thus 
$\Re\, \Phi'={\rm constant}$ on the disc. Denoting this constant by $\phi$ 
($\phi=\alpha_0$), the relation
\eqref{M} reduces to $M=2\Omega J + \phi\, Q$,
which looks like the
Smarr formula\cite{s} for an extreme Kerr-Newman black hole. For finite
$M$, the
coordinate radius
$\rho_0$ is expected to shrink to zero at the limit. Indeed, the horizon of an
extreme
Kerr-Newman black hole is 
located at $\rho=0$, $\zeta=0$ in the coordinates used here. Moreover, the
Kerr-Newman solution satisfies the conditions $f'=0$, $\Re\, \Phi'={\rm
constant}$, 
$\Im\,{\mathcal E'}=0$ and $\Im\,\Phi'=0$ on the horizon. 

A post-Newtonian expansion of the disc of charged dust solution will be
published elsewhere\cite{pm}. 
It remains a challenge to find an explicit 
analytic solution as is possible for the uncharged case ($\epsilon=0$). 

\section*{Acknowledgments}
This research was supported by the Deutsche Forschungsgemeinschaft (DFG) through
the Graduiertenkolleg 1523 ``Quantum and Gravitational Fields''.


\begin{thebibliography}{00}

\bibitem{buchdahl} H.~A. Buchdahl, {\it Phys. Rev.\/} {\bf 116}, 1027 (1959).

\bibitem{m12} R. Meinel, {\it Class. Quantum Grav.\/} {\bf 29}, 035004 (2012).

\bibitem{bw} J.~M. Bardeen and R.~V. Wagoner, {\it Astrophys. J.\/} {\bf 167},
359 (1971).

\bibitem{nm} G. Neugebauer and R. Meinel, {\it Phys. Rev. Lett.\/} {\bf 75}, 3046 (1995).

\bibitem{m02} R. Meinel, {\it Ann. Phys. (Leipzig)\/} {\bf 11}, 509 (2002).

\bibitem{rfe} R. Meinel, M. Ansorg, A. Kleinw\"achter, G. Neugebauer and D.
Petroff, {\it Relativistic Figures of 
Equilibrium\/} (Cambridge University Press, Cambridge, 2008).

\bibitem{klm} A. Kleinw\"achter, H. Labranche and R. Meinel, {\it Gen. Rel.
Grav.\/} {\bf 43}, 1469 (2011).

\bibitem{bfm} P. Breitenlohner, P. Forg\'acs and D. Maison, {\it Nucl. Phys.\/}
{\bf B 442}, 126 (1995).

\bibitem{lw} A. Lue and E.~J. Weinberg, {\it Phys. Rev.\/} {\bf D 61}, 124003 (2000).

\bibitem{akm} M. Ansorg, A. Kleinw\"achter and R. Meinel, {\it Astrophys. J.\/}
{\bf 582}, L87 (2003).

\bibitem{fha} T. Fischer, S. Horatschek and M. Ansorg, {\it Mon. Not. R.
Astron. Soc.\/} {\bf 364}, 943 (2005). 

\bibitem{lpa} H. Labranche, D. Petroff and M. Ansorg, {\it Gen. Rel.
Grav.\/} {\bf 39}, 129 (2007).

\bibitem{lemoszanchin} J.~P.~S. Lemos and V.~T. Zanchin, {\it Phys. Rev.\/} {\bf
D 81}, 124016 (2010).

\bibitem{bonnorwick} W.~B. Bonnor and S.~B.~P. Wickramasuriya, {\it Mon. Not. R.
Astron. Soc.\/} {\bf 170}, 643 (1975).

\bibitem{b98} W.~B. Bonnor, {\it Class. Quantum Grav.\/} {\bf 15}, 351 (1998).

\bibitem{b99} W.~B. Bonnor, {\it Class. Quantum Grav.\/} {\bf 16}, 4125 (1999).

\bibitem{lemoswein} J.~P.~S. Lemos and E.~J. Weinberg, {\it Phys. Rev.\/} {\bf D
69}, 104004 (2004). 

\bibitem{lz} J.~P.~S. Lemos and O.~B. Zaslavskii, {\it Phys. Rev.\/} {\bf D 76},
084030 (2007).

\bibitem{b10} W.~B. Bonnor, {\it Gen. Rel. Grav.\/} {\bf 42}, 1825 (2010).

\bibitem{mh} R. Meinel and M. H\"utten, {\it Class. Quantum Grav.\/} {\bf 28},
225010 (2011).

\bibitem{m06} R. Meinel, {\it Class. Quantum Grav.\/} {\bf 23}, 1359 (2006).

\bibitem{e} F.~J. Ernst, {\it Phys. Rev.\/} {\bf 168}, 1415 (1968).

\bibitem{s} L. Smarr, {\it Phys. Rev. Lett.\/} {\bf 30}, 71 (1973).

\bibitem{pm} S. Palenta and R. Meinel, in preparation.

\end{thebibliography}
\end{document}